# Influence of the Random Arrangement of Molecules on Energy of Vacancies Migration

## M. A. Korshunov[1]

Kirensky Institute of Physics, Siberian Division, Russian Academy of Sciences, Krasnoyarsk, 660036 Russia

**Abstract –** calculation of energy of migration for number of molecular crystals consisting of centrosymmetrical and non-centrosymmetrical molecules was carried out by using a method of atom-atom potentials. It is shown, that the potential barrier is symmetrical for crystals consisting of centrosymmetrical molecules, but not symmetrical for the mixed crystals and crystals consisting of non-centrosymmetrical molecules. Thus it is possible to influence quantity of migration energy, growing the mixed crystals with a particular arrangement of components.

As against processes of diffusion in inorganic materials the diffusion in the molecular crystals has the features. The difficulties incipient at studying of migration of molecules in low symmetrical organic crystals are connected, in particular, with non-sphericity molecules, presence of orientation oscillations, impurities, etc. One of applications of low symmetrical organic crystals is their use at record of the information. But at presence of diffusion the molecule with the changed parameters may migrate and garble the written down information. The influence of diffusion becomes especially appreciable at rise of the temperature. For reading the written down information, it is possible to use the laser. But the beam of the laser heats up a crystal. For example, beam of the He-Ne laser (30 mW) raises temperature of the molecular crystal at a place of beam passage on ~ 10 K. It may have an effect on magnification of a diffusion and longevity of the written down information. Therefore it is necessary to find requirements at which the diffusion will be diminished.

Presence of diffusion may be caused by presence of vacancies. In paper [1] using a method of a Raman Effect it is shown, that in the molecular crystals there are vacancies. Calculation of energy of vacancy migration for Naphthalene is carried out in paper [2]. It is shown, that change of a potential energy with displacement of a molecule of Naphthalene along [001] and [010] crystallographic directions aside vacancies is the symmetric concerning middle of the diagram. Therefore migration of molecules along each direction is the isotropic and consequently there is equal probability of moving forward and backward. But this isotropy may be broken in crystals consisting of non-centrosymmetrical molecules or in the mixed crystals. For checkout of it potential barriers for similar crystals were calculated.

---
[1] E-mail: mkor@iph.krasn.ru

With this purpose the following crystals were chosen: with centrosymmetrical molecules the paradiclorobenzene and the mixed crystal paradibromobenzene/paradiclorobenzene were chosen, with non-centrosymmetrical molecules the parabromochlorobenzene and the mixed crystal paradiclorobenzene/parabromochlorobenzene (this crystal is made up of both types of molecules) were chosen.

According to the X-ray data [3] and the NQR data [4] the parabromochlorobenzene, as well as paradiclorobenzene and a paradibromobenzene crystallizes in a centrosymmetrical spatial group $P2_1/a$ with two molecules in a unit cell due to statistically disordered arrangement of molecules relative to p-substituted haloids.

Interaction between molecules was described by using a method of atom-atom potentials [5]. Coefficients in the potential of interaction were used the same, as at calculations of spectrums of the lattice oscillations of a paradibromobenzene and paradiclorobenzene [6]. At calculations of structure the arrangement of vacancies was set, and there was an equilibrium configuration. By viewing migration of molecules the molecule from position (0,0,0) was step by step (0.2 A) shifted along the chosen direction aside vacancies. On each step minimization of energy was carried out.

In figures 1-4 the diagrams of change of a potential energy with displacement of investigated molecules along crystallographic directions [010] (a) and [001] (b) are shown.

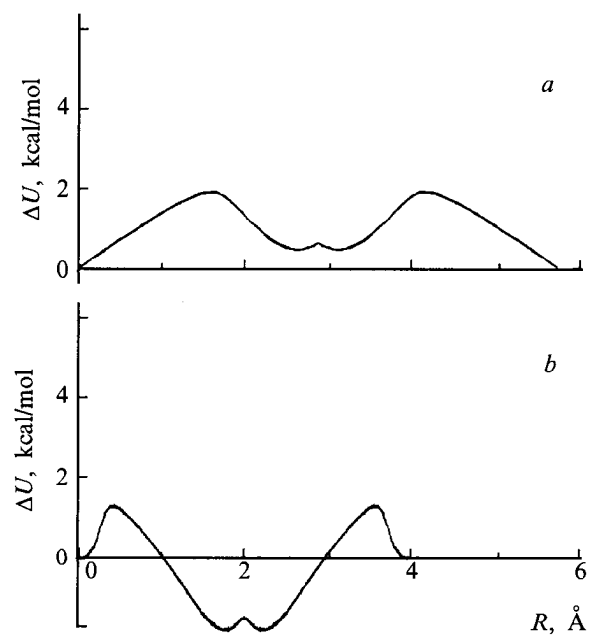

**Fig. 1.** Changes of a potential energy with displacement of molecules of paradiclorobenzene along crystallographic directions [010] (a) and [001] (b) are presented.

As we can see, for centrosymmetrical molecules of paradiclorobenzene the diagrams in figure 1 are symmetric concerning middle of the diagram and migration of a molecule both forward along each crystallographic direction, and backward is equality probability.

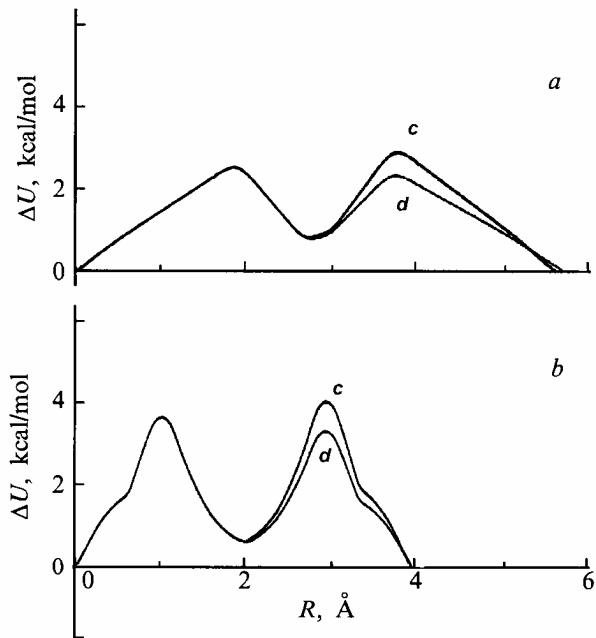

**Fig. 2.** Changes of a potential energy with displacement of an impurity (paradiclorobenzene) along crystallographic directions [010] (a) and [001] (b) in the mixed crystal a paradibromobenzene/paradiclorobenzene are presented. (c, d) are the various arrangements of molecules of components.

In figure 2 diagrams of change of a potential energy with displacement of molecules of an impurity (paradiclorobenzene) along crystallographic directions [010] (a) and [001] (b) in the mixed crystals a paradibromobenzene/paradiclorobenzene are presented. It is visible, that diagrams are the asymmetrical and migration of molecules forward and backward along crystallographic directions are not equality probability, though components of a solid solution are centrosymmetrical. In the same figure (2c and 2d) the potential barriers for various arrangements of molecules of components are shown. By selecting a particular arrangement of molecules of components it is possible to influence (reduce or increase) on amount of diffusion.

If molecules are centrosymmetrical (parabromochlorobenzene) then diagrams of change of a potential energy with displacement of a molecule aside vacancies are also the asymmetrical (figure 3).

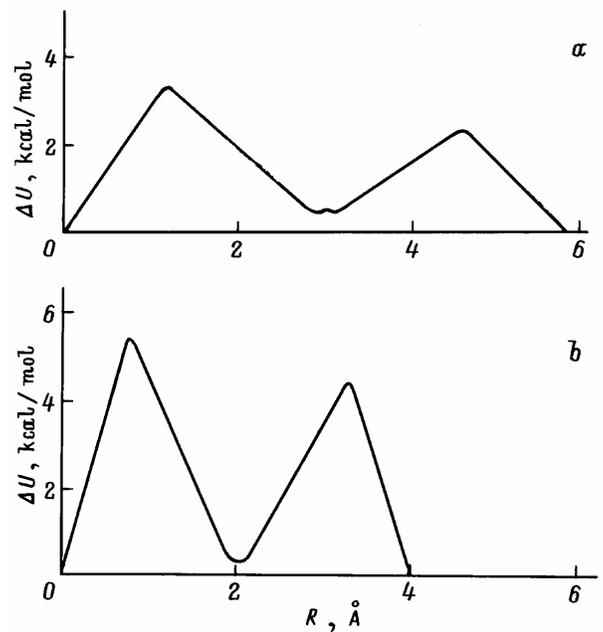

**Fig. 3.** Changes of a potential energy with displacement of molecules of parabromochlorobenzene along crystallographic directions [010] (a) and [001] (b) are presented.

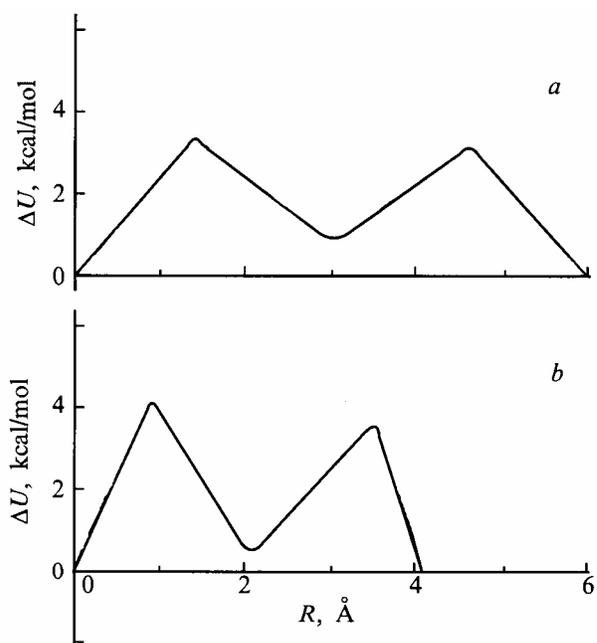

**Fig. 4.** Changes of a potential energy with displacement of an impurity (paradiclorobenzene) along crystallographic directions [010] (a) and [001] (b) in the mixed crystal a paradibromobenzene/paradiclorobenzene are presented.

Curves of change of a potential energy with displacement of a molecule of an impurity in a lattice of the mixed crystal of paradiclorobenzene with a parabromochlorobenzene along crystallographic [010] direction is shown in figure 4a and along [001] direction is shown in figure 4b. As we can see, diagrams are the asymmetrical.

Due to disordered arrangement of molecules of a parabromochlorobenzene on a lattice of the mixed crystal the environment around of a migrating molecule differs in various points of a crystal that has an effect for quantity of energy of migration. Apparently from diagrams that energy of migration may change in rather wide limits. It distinguishes the mixed crystals consisting of not centrosymmetrical molecules from crystals consisting of centrosymmetrical molecules in which change of energy of migration from cell to cell depends only on allocation of molecules of an impurity on volume of a crystal.

Thus, it is shown, that the potential barriers are symmetrical for crystals consisting of centrosymmetrical molecules, but not symmetrical for the mixed crystals and crystals consisting of non-centrosymmetrical molecules. Thus it is possible to influence quantity of energy of migration, growing the mixed crystals with a particular arrangement of components. For example by growing layered monocrystals.

## References

1. Shabanov V.F., Korshunov M.A. Fizika Tverdogo Tela (in Russian) 40, N 10, p. 1835 (1998)
2. Dautant A., Bonpunt L. Mol. Cryst. Liq. Cryst. 137, p. 221 (1986)
3. Kitaigorodskii A. I. Rentgenostrukturnyi Analiz, (X-ray Analysis), Moscow: Nauka, 1973
4. Grechishkin V. S. Nuclear Quadrupole Interactions in Solids, Moscow: Nauka, 1971
5. Kitaigorodskii A. I. Molecular Crystals, Moscow: Nauka, 1971
6. Shabanov V.F., Korshunov M.A. Fizika Tverdogo Tela (in Russian) 37, N 11, p. 3463 (1995)